\documentclass[sigconf]{acmart}

\usepackage{booktabs} % For formal tables
\usepackage{multirow}
\usepackage{rotating}
\usepackage{tabularx}
\usepackage{balance}
\usepackage{xspace}
\usepackage{pifont}
\usepackage{caption}
\usepackage{paralist}
\usepackage{hyperref}
\usepackage{booktabs}

\newcommand{\tabitem}{~~\llap{\textbullet}~~}

\newcommand{\etal}{\emph{et al.\xspace}}
\newcommand{\ie}{\emph{i.e.}, }
\newcommand{\eg}{\emph{e.g.}, }
\newcommand{\etc}{\emph{etc.\xspace}}

\setcopyright{none}
\acmConference[Submitted to MMSys'18]{Submitted to ACM Multimedia Systems Conference 2018}{June 2018}{Amsterdam, Netherlands}
\begin{document}
\title{Multi-Codec DASH Dataset}
%\subtitle{AVC, HEVC, VP9, AV1}

\begin{comment} 
\titlenote{Produces the permission block, and
  copyright information}
\subtitle{Extended Abstract}
\subtitlenote{The full version of the author's guide is available as
  \texttt{acmart.pdf} document}
\end{comment} 

\author{Anatoliy Zabrovskiy}
%\authornote{Dr.~Trovato insisted his name be first.}
\affiliation{%
  \institution{Petrozavodsk State University \& Alpen-Adria-Universit{\"a}t Klagenfurt}
  \streetaddress{Universit{\"a}tsstra{\ss}e 65-67}
  \city{Petrozavodsk}
  \postcode{9020}
  \country{Russia}
}
\email{z_anatoliy@petrsu.ru}

\author{Christian Feldmann}
%\authornote{The secretary disavows any knowledge of this author's actions.}
\affiliation{%
  \institution{Bitmovin Inc.}
  \streetaddress{301 Howard Street, Suite 1800}
  \city{San Francisco}
  \state{California}
  \country{USA}
  \postcode{94105}
}
\email{christian.feldmann@bitmovin.com}

\author{Christian Timmerer}
%\authornote{This author is the one who did all the really hard work.}
\orcid{0000-0002-0031-5243}
\affiliation{%
  \institution{Alpen-Adria-Universit{\"a}t Klagenfurt \& Bitmovin Inc.}
  \streetaddress{Universit{\"a}tsstra{\ss}e 65-67}
  \city{Klagenfurt am W{\"o}rthersee}
  \postcode{9020}
  \country{Austria}}
\email{christian.timmerer@itec.aau.at}

\begin{comment} 
\author{Valerie B\'eranger}
\affiliation{%
  \institution{Inria Paris-Rocquencourt}
  \city{Rocquencourt}
  \country{France}
}
\author{Aparna Patel}
\affiliation{%
 \institution{Rajiv Gandhi University}
 \streetaddress{Rono-Hills}
 \city{Doimukh}
 \state{Arunachal Pradesh}
 \country{India}}
\author{Huifen Chan}
\affiliation{%
  \institution{Tsinghua University}
  \streetaddress{30 Shuangqing Rd}
  \city{Haidian Qu}
  \state{Beijing Shi}
  \country{China}
}

\author{Charles Palmer}
\affiliation{%
  \institution{Palmer Research Laboratories}
  \streetaddress{8600 Datapoint Drive}
  \city{San Antonio}
  \state{Texas}
  \postcode{78229}}
\email{cpalmer@prl.com}

\author{John Smith}
\affiliation{\institution{The Th{\o}rv{\"a}ld Group}}
\email{jsmith@affiliation.org}

\author{Julius P.~Kumquat}
\affiliation{\institution{The Kumquat Consortium}}
\email{jpkumquat@consortium.net}
\end{comment} 
%\renewcommand{\shortauthors}{A. Zabrovskiy et al.}

\begin{abstract}
The number of bandwidth-hungry applications and services is constantly growing. HTTP adaptive streaming of audio-visual content accounts for the majority of today's internet traffic. Although the internet bandwidth increases also constantly, audio-visual compression technology is inevitable and we are currently facing the challenge to be confronted with multiple video codecs.

This paper proposes a multi-codec DASH dataset comprising AVC, HEVC, VP9, and AV1 in order to enable interoperability testing and streaming experiments for the efficient usage of these codecs under various conditions. We adopt state of the art encoding and packaging options and also provide basic quality metrics along with the DASH segments. Additionally, we briefly introduce a multi-codec DASH scheme and possible usage scenarios. Finally, we provide a preliminary evaluation of the encoding efficiency in the context of HTTP adaptive streaming services and applications.
\end{abstract}




\maketitle

\section{Introduction}
\label{sec:introduction}

%\todo{copy/paste from ICIP paper and needs some revision...}
Universal access to and provisioning of multimedia content is now reality. It is easy to generate, distribute, share, and consume any media content, anywhere, anytime, on any device. Interestingly, most of these services adopt a \textit{streaming} paradigm, are typically deployed over the open, unmanaged internet, and account for the majority of today's internet traffic. A major technical breakthrough and enabler was certainly HTTP Adaptive Streaming (HAS), which provides multimedia assets in multiple versions -- referred to as representations -- and chops each version into short-duration segments (\eg 2-10s) for dynamic adaptive streaming over HTTP (MPEG-DASH or just DASH)~\cite{Sodagar2011}.

Current estimations expect that the global video traffic will be about 82\% of all internet traffic by 2021~\cite{VNI}. Additionally, Nielsen's law of internet bandwidth states that the users' bandwidth grows by 50\% per year, which roughly fits data from 1983 to 2018~\cite{Nielsen}. Thus, the users' bandwidth will reach approximately 1 Gbps by 2021.

Similarly, like programs and their data expand to fill the memory available in a computer system, network applications will grow and utilize the bandwidth provided. The majority of the available bandwidth is consumed by video applications and the amount of data is further increasing due to already established and emerging applications, \eg ultra high-definition, virtual, augmented, mixed realities \etc. Therefore, video compression is inevitable and every new major video codec has increased coding efficiency significantly compared to its predecessor\footnote{\url{http://blog.chiariglione.org/a-crisis-the-causes-and-a-solution/}, accessed Mar 3, 2018.}. Currently, we are in a situation where we can choose from multiple video codecs such as Advanced Video Coding (AVC)~\cite{AVCOverview}, High Efficiency Video Coding (HEVC)~\cite{HEVCOverview}, VP9~\cite{VP9Overview}, and AOMedia Video 1 (AV1)~\cite{AV1Software}. However, not every kind of end user device (\ie ranging from smart phone to smart TV) is supporting all possible codecs -- except maybe AVC, which has significant lower coding efficiency than others -- and, thus, there is a need to support multiple codecs while we are in the transition to a new commonly agreed universal video codec supported across all end user devices (if this ever will happen). Consequently, in this paper we propose a \textbf{multi-codec DASH dataset} comprising all aforementioned video codecs including state-of-the-art bitrate/resolution configurations and quality metrics (\ie PSNR, SSIM). The goal of such a dataset is to enable
\begin{inparaenum}[\itshape (i)\upshape]
\item interoperability testing and
\item experimenting with different adaptation strategies 
\end{inparaenum}
of DASH clients supporting multiple video codecs. Note that the AV1 syntax and tools are not yet finalized at the time of writing of this paper and streaming is currently working with Firefox Nightly\footnote{\url{https://demo.bitmovin.com/public/firefox/av1/}, accessed: Mar 3, 2018.}. We will update AV1 encoded assets of the dataset as soon as the AV1 syntax and tools haven been frozen and officially released.

Most importantly, our dataset is available here:
\begin{center}
\textbf{\url{http://www.itec.aau.at/ftp/datasets/mmsys18/}}
\end{center}

The remainder of this paper is structured as follows. Section~\ref{sec:related} highlights related work. The methodology and the actual dataset is described in Section~\ref{sec:method}. Some preliminary evaluation results are provided in Section~\ref{sec:results}. Section~\ref{sec:conclusions} concludes the paper and highlights future work.
\section{Related work}
\label{sec:related}

In the past we have witnessed many DASH datasets, which are briefly highlighted in this section. The first DASH dataset was released by Lederer~\etal~\cite{Lederer:2012} and comprises various genres (\ie animation, sport, movie), encoded using up to 20 representations (up to 1080p resolution), and different segment lengths (\ie 1, 2, 4, 6, 10, and 15 seconds). Additionally, for some representations per frame PSNR values are provided. Initial evaluations of the dataset provide recommendations for an optimal segment length based on the coding efficiency (\ie 4s) and the influence of enabled versus disabled persistent connections.

A distributed DASH dataset has been released by Lederer~\etal~\cite{Lederer:2013}, which distributes the dataset across multiple locations and utilizes multiple \textit{BaseURL} elements within the media presentation description (MPD). It can be used to simulate different content distribution network (CDN) locations and bitstream switching across multiple CDNs.

Le Feuvre~\etal~\cite{LeFeuvre:2014:UHD} provide an ultra high definition (UHD) HEVC DASH dataset targeting UHD services (\ie resolutions up to 3840x2160, framerate up to 60 fps, and up to 10 bpp) using HEVC, which is the major difference compared to previously proposed datasets. Kreuzberger~\etal~\cite{Kreuzberger:2015:SVC} provides a DASH dataset focusing on scalable video coding (SVC) and experimenting with in-network adaptation in named data networks and information-centric networking, respectively. Unfortunately, support for SVC in end user devices is still very limited. Finally, Quinlan~\etal~\cite{Quinlan:2016} propose a dataset comprising AVC and HEVC for the evaluation of DASH systems and comes closest to what we propose in this paper, namely a multi-codec DASH dataset, which is described and evaluated in the following sections.
\section{Methodology and Multi-Codec DASH Dataset Description}
\label{sec:method}
%\subsection{Content}

This section describes the methodology and conception of the multi-codec DASH dataset.

\subsection{Content Sequences}
%\todo{See comment in \LaTeX} 
The goal of this paper is to cover a common situation with a dataset comprising of different types of video content. In order to achieve this, we take into account the spatial information (SI) and temporal information (TI)~\cite{itu910.2008, SITI:software} of the video sequences. Therefore, our dataset comprises video sequences with minor movements, such as moving head on a fixed black background and sequences with significant movements, such as riding jockeys. It is worth noting that some test sequences used in this paper are based on a subset of the dynamic adaptive streaming over HTTP dataset~\cite{Lederer:2012}, which has been introduced to experiment with HAS applications and services. In particular, we selected three long sequences: Big Buck Bunny (BBB), Sintel, and Tears of Steel (TOS). The former two are computer-animated short movies and the latter is a mix of natural scenes, computer-generated content, and combinations thereof, which makes it very interesting for our dataset as it represents a wide range of different use cases. For these sequences we used an excerpt of 60s, which starts with the $61^{st}$ second.
Additionally, we selected five short sequences (\ie Beauty, HoneyBee, Jockey, ReadySetGo, YachtRide)\footnote{http://ultravideo.cs.tut.fi/\#testsequences, accessed Mar 3, 2018.} with a duration of 20s.
%each were created by the laboratory of Pervasive Computing\footnote{http://www.tut.fi/en/research/research-fields/pervasive-computing/index.htm} (Tampere University of Technology).
Finally, we selected two 20s sequences from Netflix test videos (\ie DrivingPOV and WindAndNature)\footnote{https://media.xiph.org/video/derf, accessed Mar 3, 2018.}. The main reason for adopting rather short sequences (\ie 20s and 60s) is due to the current encoding performance of AV1, which is not yet optimized for speed or parallel processing. We expect to update the dataset once AV1 has been optimized in this respect.

\begin{table*}[pt!]
\centering
\small
%\label{tab:originalfilechars}
\caption{Original video file characteristics.}\label{tab:originalfilechars}\vspace{4pt}
\begin{tabular}{ccccccccc} 
\hline 
\textbf{Characteristic} & \textbf{Genre} & \textbf{Creator}  & \textbf{Frame rate} & \textbf{Resolution} & \textbf{File format} & \textbf{File duration} & \textbf{Sequence duration} \\
\hline
        \small{BBB} & \small{Animation} & \small{Blender Foundation} & \small{30 fps} & \small{3840x2160} & \small{mp4} & \small{634 sec.} & \small{60 sec.} \\
        \small{Beauty} & \small{Moving head} & \small{TUT, Finland} & \small{30 fps} & \small{3840x2160} & \small{y4m} & \small{20 sec.} & \small{20 sec.} \\
        \small{DrivingPOV} & \small{Moving cars} & \small{Netflix, Inc.} & \small{60 fps} & \small{4096x2160} & \small{y4m} & \small{20 sec.} & \small{20 sec.} \\     
        \small{HoneyBee} & \small{Nature} & \small{TUT, Finland} & \small{30 fps} & \small{3840x2160} & \small{y4m} & \small{20 sec.} & \small{20 sec.} \\
        \small{Jockey} & \small{Moving jockey} & \small{TUT, Finland} & \small{30 fps} & \small{3840x2160} & \small{y4m} & \small{20 sec.} & \small{20 sec.} \\ 
        \small{ReadySetGo} & \small{Moving horses} & \small{TUT, Finland} & \small{30 fps} & \small{3840x2160} & \small{y4m} & \small{20 sec.} & \small{20 sec.} \\ 
        \small{Sintel} & \small{Animation} & \small{Blender Foundation} & \small{24 fps} & \small{4096x1744} & \small{y4m} & \small{888 sec.} & \small{60 sec.} \\
		\small{TOS} & \small{Composite} & \small{Blender Foundation} & \small{24 fps} & \small{4096x1714} & \small{y4m} & \small{734 sec.} & \small{60 sec.} \\
        \small{WindAndNature} & \small{Rotating wind vanes} & \small{Netflix, Inc.} &  \small{60 fps} & \small{4096x2160} & \small{y4m} & \small{20 sec.} & \small{20 sec.} \\       
        \small{YachtRide} & \small{Moving yacht} & \small{TUT, Finland} & \small{30 fps} & \small{3840x2160} & \small{y4m} & \small{20 sec.} & \small{20 sec.} \\     
		\hline
\end{tabular}
\end{table*}

\begin{figure}[pt!]
\centering
\includegraphics[scale=0.55]{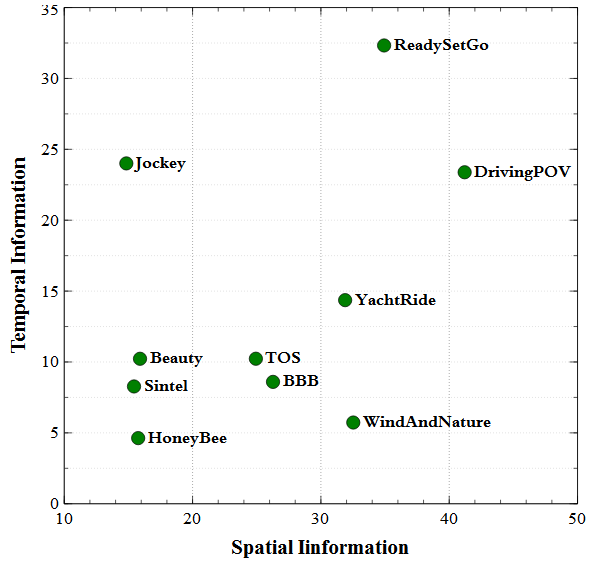}
\caption{Average spatial information (SI) and temporal information (TI) for dataset sequences.}
\label{fig:avg-siti}
\end{figure}

The temporal and spatial information for all sequences are shown in Fig.~\ref{fig:avg-siti}. The figure shows that the chosen test set covers a wide range of different sequence types and genres. The parameters of all sequences used in our dataset are shown in Table~\ref{tab:originalfilechars}.
%Short video sequences were chosen in such a way that they relate to different values of spatial and temporal characteristics.

% The paragraph above gives the impression that something is missing or is suboptimal. Please clearly state in a positive way why these sequences has been selected. It is a bit hidden in the last sentence mentioning SI and TI which needs to be more emphasized.

\begin{table}[pt!]
\centering
\normalsize
\caption{Bitrate ladder (bitrate/resolution pairs) of the multi-codec DASH dataset.}
\label{tab:bitrateladder}
\begin{tabular}{c@{\hspace{0.5em}}c@{\hspace{0.5em}}c@{\hspace{2em}}c@{\hspace{0.5em}}c@{\hspace{1em}}c}
        \hline
        \textbf{\#} & \textbf{Bitrate} & \textbf{Resolution} & \textbf{\#} &\textbf{Bitrate} & \textbf{Resolution} \\ 
        \hline 
        1 & 100 & 256x144 & 11 & 4300 & 1920x1080 \\
		2 & 200 & 320x180 & 12 & 5800 & 1920x1080 \\
		3 & 240 & 384x216 & 13 & 6500 & 2560x1440 \\
		4 & 375 & 384x216 & 14 & 7000 & 2560x1440\\
		5 & 550 & 512x288 & 15 & 7500 & 2560x1440 \\
		6 & 750 & 640x360 & 16 & 8000 & 3840x2160\\
		7 & 1000 & 768x432 & 17 & 12000 & 3840x2160 \\
		8 & 1500 & 1024x576 & 18 & 17000 & 3840x2160 \\
		9 & 2300 & 1280x720 & 19 & 20000 & 3840x2160 \\
		10 & 3000 & 1280x720 \\        
		\hline
\end{tabular}
\end{table}

The main focus of our dataset is related to HAS services and, thus, we adopted the bitrate/resolution pairs -- referred to as \textit{bitrate ladder} -- as shown in Table~\ref{tab:bitrateladder}. This selection is based on existing datasets proposed in the literature~\cite{Lederer:2012, Lederer:2013, LeFeuvre:2014:UHD, Quinlan:2016} as well as industry best-practices and guidelines~\cite{NetflixBitrates:2016, Apple:2017, Twitch, Ustream, YouTube}. The bitrate ladder covers a wide range of bitrate/resolutions including support for ultra high-definition services. Finally, the segment length is an important parameter in HAS services as typically each segment starts with a random access point to enable dynamic switching to other representations at segment boundaries. Therefore, we adopted 2s and 4s segment lengths for this dataset. The latter shows the best trade-off with respect to streaming performance and coding efficiency~\cite{Lederer:2012} and is also adopted within commercial deployments. The former is still also used in commercial deployments and confirms the trend towards low-latency requirements~\cite{Bouzakaria2014}.

\subsection{Encoding, Packetization, and DASH Parameters}

The actual encoding and generation of the HAS segments is conducted taking pre-segmented sequences corresponding to the target segment length as input in order to enable parallel processing, \eg within cloud environments. \textit{ffmpeg} is used to generate these pre-segmented sequences with 2s and 4s segment time, respectively. The encoding for AVC, HEVC, and VP9 is performed utilizing \textit{ffmpeg} and, thus, \textit{libx264}, \textit{libx265}, and \textit{libvpx-vp9} are used as follows: {\footnotesize\texttt{{ffmpeg -y -i \{input.y4m\} -r 24 -vf scale=\{WxH\} format=yuv420p -c:v \{libx264,libx265,libvpx-vp9\}  -preset slow -pass \{1|2\} -b \{bitrate\} \{out.mp4|out.webm\}}}}. For AV1 we used the AOM reference software as follows \footnote{The software is freely available at \cite{AV1Software}. The used version is v0.1.0-7691-g84dc6e9 (Git-Hash 84dc6e97cb6bfaee4185e63ac78dcd5e080f378c).}: {\footnotesize\texttt{{aomenc \{input.y4m\} -v -\phantom{}-good -\phantom{}-cpu-used=2 -\phantom{}-width=\{W\} -\phantom{}-height=\{H\} -\phantom{}-target-bitrate=\{bitrate\} -\phantom{}-psnr -o \{out.webm\}}}}. The goal for the established codecs (\ie AVC, HEVC, and VP9) was to choose settings, which use a similar trade-off between performance and speed as for the AV1 encoder with \texttt{cpu\_used} set to 2. This includes the motion estimation algorithm as well as other estimation algorithms. Additionally, we used two-pass encoding for AVC, HEVC, and VP9 as this is default in AV1.

For the packetization we adopted what is provided by \textit{ffmpeg} and \textit{aomenc} by default as we use pre-segmented sequences corresponding to the target segment length as input. That is, for AVC and HEVC we provide MP4 segments (\ie MPEG-4 Part 15 Carriage of NAL unit structured video in the ISO Base Media File Format) and for VP9 and AV1 we use the WebM container format.

Finally, the DASH MPD is created in a way that representations of each video codec are within a single \textit{AdaptationSet}, which allows seamless switching within a specific codec but also facilitates switching across codecs if supported by the DASH client and device, respectively. Each segment is available as a single physical file, which allows for maximum flexibility in adaptive streaming experimental setups. The DASH \textit{subsegment} approach can be realized by concatenating segments of a single representation, but this requires to update the MPD and/or adding a segment index box (sidx) to the concatenated representation. 

%\todo{DASH parameters, how does the MPD look like? I'd propose a AdaptationSet per codec...}

\subsection{Multi-Codec DASH Scheme}

% I think it's better if you explain Fig.~\ref{fig:multi-codec} first, explain all submodules, then also Table~\ref{tab:client-server} etc. and then present a possible use case Fig.~\ref{fig:codecs}

\begin{figure}[pt!]
\centering
\includegraphics[scale=0.45]{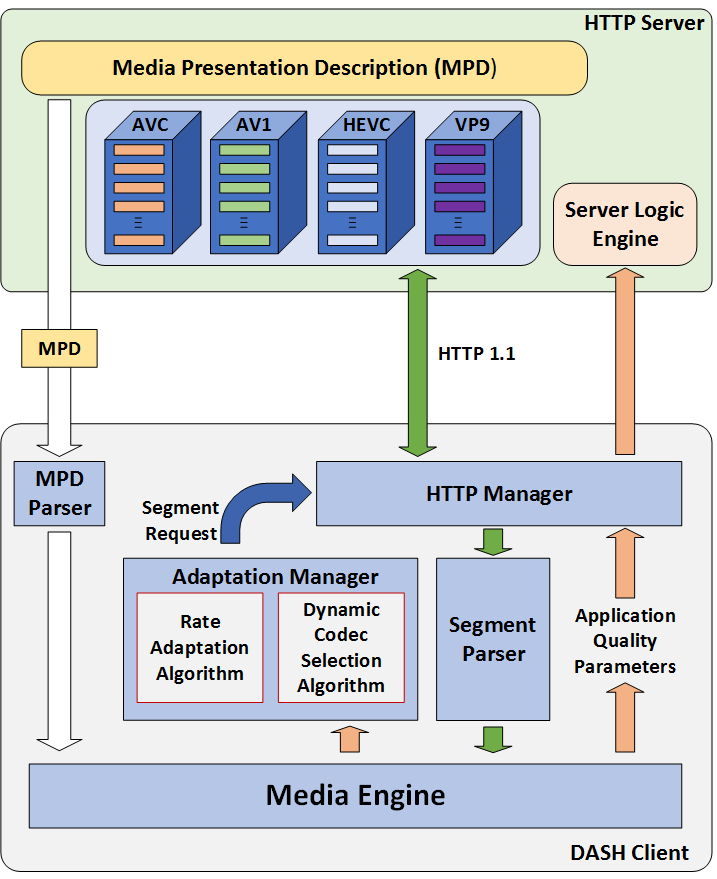}
\caption{Multi-codec DASH scheme.}
\label{fig:multi-codec}
\end{figure}

In this section we present our multi-codec DASH scheme and its potential to be adopted for HAS is reviewed. Fig.~\ref{fig:multi-codec} illustrates the multi-codec DASH scheme comprising a traditional \textit{HTTP Server} and a \textit{DASH Client}. The HTTP server infrastructure is used to store the \textit{Media Presentation Description (MPD)} and media segments encoded with different codecs. The MPD contains information for the DASH client for adaptive streaming of the content. The \textit{Server Logic Engine} of HTTP server is responsible for obtaining, accumulating, and analyzing information received from the DASH client. That is, it analyzes client requests and application quality parameters, automatically encodes new or removes unused segments and updates MPDs. Note that the functionality of the server logic engine could be also realized on a different instance.

The DASH client consist of the following modules:
\begin{itemize}
\item \textbf{HTTP Manager} --- responsible for all HTTP requests and responses between the HTTP server and the DASH client.
\item \textbf{MPD Parser} --- parses MPD and checks that all input data in MPD has been provided appropriately.
\item \textbf{Segment Parser} --- handles received segments and encapsulated coding formats.
\item \textbf{Media Engine} --- responsible for decoding and rendering media presentations. It also sends application quality parameters (video startup time, download video bitrate, video buffer length, etc.) to the server logic engine.
\item \textbf{Adaptation Manager} --- includes two algorithms, as shown in the Fig.~\ref{fig:multi-codec}. One is responsible for the \textit{rate adaptation} and one for the \textit{dynamic codec selection}.
\end{itemize}

\begin{table}[pt!]
\centering
\small
%\label{tab:originalfilechars}
\caption{Server and client roles in a multi-codec DASH scheme.}\label{tab:client-server}\vspace{4pt}
    \begin{tabular}{ | p{3.8cm} | p{3.8cm} |}
    \hline
    \textbf{Server-side roles} & \textbf{Client-side roles} \\ \hline
    \tabitem Compute quality metrics of encoded segments. & \tabitem Use quality metrics to decide which segments to request.\\
    \tabitem Deliver quality metrics to the client. & \tabitem Send application quality parameters to the server. \\ 
    \tabitem Analyze clients requests and application quality parameters.  & \tabitem Switch between segments encoded with different codecs.   \\ 
    \tabitem Automatically encode new or remove unused segments; update MPDs. & \tabitem Switch between representations of the video stream.   \\  \hline
    \end{tabular}
\end{table}

\begin{figure}[pt!]
\centering
\includegraphics[scale=0.48]{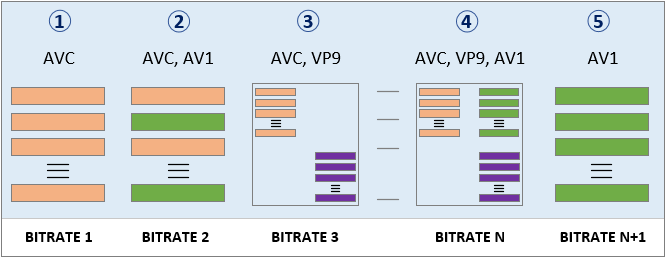}
\caption{Multi-codec DASH dataset use cases.}
\label{fig:codecs}
\end{figure}

The server and client roles include tasks and operations summarized in Table~\ref{tab:client-server}. We assume that some decisions regarding the coding of new or deleting of stored representation can be made by the server logic engine automatically using the tools of machine learning. Moreover, both algorithms of the adaptation manager also can work with the use of machine learning techniques~\cite{Mao:2017:NAV:3098822.3098843}. The best streaming scheme could be achieved through a trade-off between codec bandwidth requirements and desired quality. Some situations that can lead to the choice of multi-codec scheme as the delivery strategy for adaptive streaming services are shown in Fig.~\ref{fig:codecs} and summarized below: 
%\todo{AZ, please check the text below; it's pretty generic and vague, maybe we can be a bit more concrete}
\begin{itemize}
\item Low bitrate streams are first encoded using fast approach, such as AVC (\eg \ding{172} of Fig.~\ref{fig:codecs}). That allows a streaming system to provide video content to viewers in a short time. 
\item Some streams can include segments encoded with different codecs. For example, segments with a large number of motions are encoded by a more efficient codec (\eg \ding{173} of Fig.~\ref{fig:codecs}).
\item Some streams can be encoded in several adaptation sets of segments by two or more codecs (\eg \ding{174} and \ding{175} of Fig.~\ref{fig:codecs}). This can happen with bitrate representations which are most frequently transmitted.
\item Very high bitrates are encoded by the most efficient codecs. This takes a long time but significantly saves bandwidth (\eg \ding{176} of Fig.~\ref{fig:codecs}). 
\end{itemize}

It should be noted that saving a media stream on the HTTP server encoded with different codecs will allow its viewing on different devices and in all popular end user devices including web browsers. A more detailed analysis of the possible implementation schemes for the adaptation manager is out of scope for this dataset paper.

\section{Preliminary Evaluation Results}
\label{sec:results}

%\todo{present some results on PSNR and SSIM incl. BD -- well, basically we have up to 6 pages}

In this section, we present a preliminary coding performance evaluation of AV1 compared to AVC, HEVC, and VP9 in the context of HAS. Therefore, we measured the weighted Peak Signal-to-Noise Ratio (wPSNR) for the luminance (Y) and chrominance (UV) components according to~\cite{Ohm2012}:

\begin{align}
  \textsf{$wPSNR$} &= {{6{\cdot}PSNR_Y + PSNR_U + PSNR_V \over 8}}
  \label{eq:wpsnr}
\end{align}

PSNR and bitrate is averaged over all frames in each sequence and the Bj{\o}ntegaard-Delta bit-rate (BD-rate)~\cite{Bjontegaard:2001} is calculated from these values over the entire bitrate ladder as depicted in Table~\ref{tab:AV1-vs-others}. It can be seen that AV1 is able to outperform all the other codecs. The average gains over AVC are around 48\%, 17\% compared to HEVC, and 13\% compared to VP9. It can also be seen that these gains vary strongly and depend on the individual dataset sequences. For HEVC, the gains range from close to zero up to 31\%. It should be noted that the sequence BBB represents a special situation, \ie the sequence BBB is a computer-animated clip with only very little motion. Thus, this sequence is particularly easy to compress, which can be seen in the very high PSNR values. As a result, the x265 encoder did not adhere to the bitrate, which was set by the bitrate ladder and, thus, used a lower bitrate. Therefore, the HEVC results for BBB should be interpreted with caution. In particular, the chosen bitrate ladder for the BBB sequence is not optimal and could be improved by, \eg an per title encoding approach~\cite{PerTitleEncoding}.

In Table~\ref{tab:AV1_2160p-vs-others}, the same comparison is performed specifically for the four bitrate points (\ie 8, 12, 17, and 20 Mbps) at the highest resolution of 3840x2160 (see Table~\ref{tab:bitrateladder}). It can be seen that specifically for this high resolution and high bitrates, the coding gain of AV1 compared to the other codecs is even higher with an average of roughly 58\% compared to AVC, 44\% compared to HEVC, and 27\% compared to VP9. It should be noted, that for some of the sequences the PSNR values are quite high and the results are specific to the selected bitrate ladder. Also for this reason some of the BD-rate calculations comparing AV1 to AVC fail because there is no overlap on the PSNR axis between the two rate-distortion curves in this setting. For the sequence BBB, no encoding at this high resolution was performed. Please note that this could be further optimized using a more advanced bitrate ladder adaption technique as mentioned above.

The raw values of PSNR and also SSIM are part of the dataset and available for further analysis. We did not include other metrics such as VMAF as VMAF has been designed and trained for resolutions up to full HD whereas our bitrate ladder also includes UHD resolutions.

\begin {table}[pt]
\centering
\normalsize
\caption{BD-BR for segments of AV1 compared to AVC, HEVC, and VP9 over the entire bitrate ladder.}
\label{tab:AV1-vs-others}
\begin{tabular}{ccccc}
\hline
%\cline{1-4}
\textbf{Sequence} & \multicolumn{3}{c}{\textbf{BD-rate}} \\
         & \textbf{AVC} & \textbf{HEVC} & \textbf{VP9} \\
\hline
BBB 2 sec. & -1.59 \% & 12.84 \% & 0.63 \%  \\
BBB 4 sec. & -5.20 \% & -0.19 \% & -2.95 \%  \\

Beauty 2 sec. & -85.51 \% & -14.46 \% & -37.86 \%  \\
Beauty 4 sec. & -62.12 \% & -3.05 \% & 1.04 \%  \\

DrivingPOV 2 sec. & -64.26 \% & -11.52 \% & -9.97 \%  \\
DrivingPOV 4 sec. & -64.15 \% & -5.29 \% & -0.87 \% \\

HoneyBee 2 sec. & -37.11 \% & -30.53 \% & -24.99 \%  \\
HoneyBee 4 sec. & -42.39 \% & -29.48 \% & -20.91 \%  \\

Jockey 2 sec. & -62.44 \% & -24.35 \% & -14.99 \%  \\
Jockey 4 sec. & -60.79 \% & -21.83 \% & -4.74 \%  \\

ReadySetGo 2 sec. & -49.90 \% & -18.57 \% & -12.76 \%  \\
ReadySetGo 4 sec. & -54.23 \% & -21.30 \% & -12.17 \% \\

Sintel 2 sec. & -26.71 \% & -14.47 \% & -6.05 \%  \\
Sintel 4 sec. & -24.45 \% & -13.64 \% & -6.97 \%  \\

TOS 2 sec. & -22.35 \% & -0.80 \% & -4.17 \%  \\
TOS 4 sec. & -23.36 \% & -14.09 \% & -7.84 \%  \\

WindAndNature 2 sec. & -33.22 \% & -21.32 \% & -8.96 \%  \\
WindAndNature 4 sec. & -43.55 \% & -23.72 \% & -13.01 \%  \\

YachtRide 2 sec. & -52.83 \% & -18.11 \% & -31.86 \%  \\
YachtRide 4 sec. & -55.86 \% & -20.90 \% & -21.94 \%  \\
\hline
\textcolor{black}{\textbf{Average value}} & \textcolor{black}{\textbf{-48.07 \%}} & \textcolor{black}{\textbf{-17.08 \%}} & \textcolor{black}{\textbf{-13.28 \%}}  \\

\hline
\end{tabular}
\end{table}

\begin {table}[pt]
\centering
\normalsize
\caption{BD-BR for segments of AV1 at 3840x2160 compared to AVC, HEVC, and VP9.}
\label{tab:AV1_2160p-vs-others}
\begin{tabular}{ccccc}
\hline
%\cline{1-4}
\textbf{Sequence} & \multicolumn{3}{c}{\textbf{BD-rate}} \\
         & \textbf{AVC} & \textbf{HEVC} & \textbf{VP9} \\
\hline
Beauty 2 sec. & --- & -50.20 \% & -23.42 \%  \\
Beauty 4 sec. & --- & -46.54 \% & -6.71 \%  \\

DrivingPOV 2 sec. & -54.09 \% & -23.82 \% & -20.91 \%  \\
DrivingPOV 4 sec. & -49.84 \% & -15.82 \% & -13.71 \% \\

HoneyBee 2 sec. & --- & -51.09 \% & -41.32 \%  \\
HoneyBee 4 sec. & --- & -55.91 \% & -48.31 \%  \\

Jockey 2 sec. & --- & -60.67 \% & -22.42 \%  \\
Jockey 4 sec. & --- & -60.34 \% & -22.84 \%  \\

ReadySetGo 2 sec. & -63.60 \% & -39.25 \% & -27.55 \%  \\
ReadySetGo 4 sec. & -62.99 \% & -35.87 \% & -25.30 \% \\

Sintel 2 sec. & -61.77 \% & -42.20 \% & -12.40 \%  \\
Sintel 4 sec. & --- & -39.38 \% & -12.30 \%  \\

TOS 2 sec. & -67.14 \% & -44.86 \% & -23.89 \%  \\
TOS 4 sec. & --- & -42.71 \% & -29.34 \%  \\

WindAndNature 2 sec. & --- & -59.04 \% & -42.06 \%  \\
WindAndNature 4 sec. & --- & -58.79 \% & -47.63 \%  \\

YachtRide 2 sec. & -52.76 \% & -31.86 \% & -28.59 \%  \\
YachtRide 4 sec. & -52.93 \% & -31.92 \% & -30.44 \%  \\
\hline
\textcolor{black}{\textbf{Average value}} & \textcolor{black}{\textbf{-58.14 \%}} & \textcolor{black}{\textbf{-43.90 \%}} & \textcolor{black}{\textbf{-26.62 \%}}  \\
\hline
\end{tabular}
\end{table}

\section{Conclusions}
\label{sec:conclusions}

It this paper, we present a multi-codec DASH dataset comprising multiple state-of-the-art as well as emerging video codecs, \ie AVC, HEVC, VP9, and AV1. We expect that DASH services will have to deal with such situations and, thus, we offer this dataset for researchers and practitioners in the field of HAS. The dataset has been designed based on existing DASH datasets and industry guidelines to reflect current and future requirements of DASH services. It can be used for interoperability testing and for experimenting with adaptation strategies of DASH clients supporting multiple video codecs. Some of these strategies have been briefly highlighted in this paper but we expect users of the dataset to investigate them in more detail.

We also present a preliminary evaluation of the encoding performance of AV1 compared to AVC, HEVC, and VP9. Interestingly, AV1 outperforms AVC by 48\%, HEVC by 17\%, and VP9 by 13\% over the entire bitrate ladder. Performance gains for higher bitrates and a higher resolution are even higher. Please note that this evaluation primarily targets HAS services and should not be used as a general purpose codec evaluation. 

Future work includes
\begin{inparaenum}[\itshape (i)\upshape]
\item -- most importantly -- updating the dataset once AV1 syntax is frozen and preparing for AV2, eventually the successor or AV1;
\item adding support for MP4/ISOBMFF bindings of AV1, which are under development; and
\item adding support for the successor of HEVC, which is currently developed by the Joint Video Exploration Team (JVET) of ITU-T VCEG (Q6/16) and ISO/IEC MPEG (JTC 1/SC 29/WG 11).
\end{inparaenum}

%Conclusions go here... well, basic usage as other, similar datasets in this area; main focus is multiple codecs which are supported in today's infrastructure environments.

%Highlight some results from the codec comparison (summary only).

%Future work, add support for future video coding (aka H.266 or whatever will be the name of the successor of HEVC). Also AV2 is coming and there's also AVS (important for China but less relevant for the rest).

%Importance of machine learning technologies for adaptive video streaming.

%Currently, AV1 is packetized with webm but isobmff bindings are defined also.

Finally, our dataset is available here:
\begin{center}
\textbf{\url{http://www.itec.aau.at/ftp/datasets/mmsys18/}}
\end{center}

\begin{acks}
This work was supported in part by the Austrian Research Promotion Agency (FFG) under the Next Generation Video Streaming project "PROMETHEUS".
\end{acks}

\balance
\bibliographystyle{ACM-Reference-Format}
\bibliography{sample-bibliography}

\end{document}